\documentclass[]{article}
\usepackage[]{amsmath}
\usepackage[]{amssymb}
\usepackage[]{amsthm}
\usepackage[a4paper]{geometry}
\usepackage{graphicx}
\usepackage[numbers]{natbib}
\usepackage{url}
\usepackage{authblk}
\usepackage{setspace}
\usepackage{xcolor}
\usepackage{floatpag}
\usepackage{pdflscape}
\usepackage{rotating}
\usepackage{enumitem}
\usepackage[aboveskip=1pt,labelfont=bf,labelsep=period,justification=raggedright,singlelinecheck=off]{caption}

\begin{document}
	
	\title{Multivariable Mendelian randomization with weak instruments: a comparison of Bayesian and frequentist methods}
	\author[1]{Andrew J. Grant\thanks{Corresponding author. Email address: andrew.grant1@sydney.edu.au}}
	\author[2]{Ashish Patel}
	\author[2,3]{Stephen Burgess}
	\affil[1]{\normalsize Sydney School of Public Health, University of Sydney, Sydney, Australia}
	\affil[2]{\normalsize MRC Biostatistics Unit, University of Cambridge, Cambridge, UK}
	\affil[3]{\normalsize Cardiovascular Epidemiology Unit, University of Cambridge, Cambridge, UK}
	\date{}
	\maketitle
	
	\onehalfspacing
	
	\begin{abstract}
		Weak instruments are a well known limitation for valid causal inference in Mendelian randomization studies. In the single exposure setting, weak instrument bias can be mitigated by selecting genetic instruments which are strongly associated with the exposure according to p-value and/or F-statistic thresholds. However, in the multi-exposure setting, genetic instruments may be strongly associated with an exposure but weakly associated with it conditional on all other exposures in the analysis. It is therefore more difficult to guarantee conditionally strong instruments in multivariable Mendelian randomization. Weak instrument bias can be mitigated using modelling approaches, however there are fewer methods for doing this in the multivariable case compared with the single exposure case. In this paper, we consider a method for mitigating weak instrument bias in multivariable Mendelian randomization using a Bayesian framework: MVMR-Pony. We compare this method with existing frequentist methods. We show using simulation studies that the MVMR-Pony method outperforms the frequentist approaches with respect to bias, coverage, type I error rates, and power, across settings where weak instrument bias arises due to correlated genetic effects, measurement error, and mediation.
	\end{abstract}
	
	\section*{Introduction}
		Mendelian randomization is a technique in genetic epidemiology which uses genetic variants as instrumental variables to infer causal effects of an exposure on an outcome \cite{Lawlor2008}. For a genetic variant to be considered a valid instrumental variable, three assumptions are required, which are: the variant is associated with the exposure; the variant is not associated with the outcome via a confounding pathway; and the variant does not affect the outcome other than potentially via its effect on the exposure \cite{Greenland2000}. If a valid genetic instrument is associated with the outcome, this implies a causal effect of the exposure on the outcome. Furthermore, making an additional homogeneity assumption, consistent estimators of the causal effect can be obtained \cite{Hernan2006}. Popular approaches for Mendelian randomization estimate causal effects using summary statistics from genome-wide association studies (GWAS), precluding the need for access to individual level data. They are also able to incorporate multiple genetic instruments which can increase the power to detect a causal effect \cite{Burgess2013IVW}.
		
		Multivariable Mendelian randomization is an extension to the Mendelian randomization framework which is able to consider multiple exposure variables simultaneously \cite{BurgessThompson2015mvmr}. A genetic variant is a valid instrumental variable in the multivariable setting if it: associates with at least one of the exposures; is not associated with the outcome via a confounding pathway; and does not affect the outcome other than potentially via its effect on one or more of the exposures. Multivariable Mendelian randomization can be used to account for pleiotropic pathways between the genetic variant and the outcome through measured covariates. Increasingly, multivariable Mendelian randomization is being used in applications such as risk factor selection \cite{Zuber2021}, mediation analysis \cite{Carter2021}, and studies of disease progression \cite{Cai2024}.
		
		The essence of the instrumental variables framework is that, when the assumptions are met, an instrument can be considered to be an unconfounded proxy for the exposure. However, in practice with a finite sample size, there will always be some level of sample correlation between the instrument and confounders of the exposure-outcome relationship(s) due to random variation. If this sample correlation is similar, or greater, in magnitude to the correlation between the instrument and the exposure(s), then the instrument will no longer act as an unconfounded proxy. The result is that causal effect estimates may be biased, despite the use of valid instruments. This bias is known as weak instrument bias, as it arises when the correlation between the instrument and exposure(s) is relatively low. With an infinite sample size, this bias will not occur, as then the sample correlation between a valid instrument and any exposure-outcome confounders will be zero.
		
		Conventionally, the F-statistic obtained from a regression of an exposure on the instruments is used as a measure of instrument strength for that exposure, with a value below 10 a common threshold for defining a weak instrument \cite{StaigerStock1997}. With summary level data, the F-statistic can be estimated as the mean of the squared t-statistics corresponding to each instrument.
		
		In multivariable Mendelian randomization, it is possible for an instrument to be strongly associated with an exposure, but weakly associated with that exposure conditional on the other exposures in the analysis. The more relevant measure of instrument strength in the multivariable setting is therefore not the F-statistic, but the conditional F-statistic. The conditional F-statistic can be defined as the F-statistic obtained from a regression on the instruments of the fitted values from a regression of the exposure of interest on all the other exposures. That is, it reflects the proportion of the variation in the exposure of interest that is independently explained by the genetic variants. If individual level data is not available, Sanderson et al. \cite{Sanderson2021} and Patel et al. \cite{Patel2023} have described methods for estimating conditional F-statistics from GWAS summary statistics.
		
		Two-sample Mendelian randomization refers to the scenario where the summary statistics for the exposures and the outcome are taken in independent samples from the same population \citep{PierceBurgess2013}. There are a variety of methods for estimating causal effects in the two-sample framework \citep{Sanderson2022}. When all genetic variants are valid instrumental variables, the inverse-variance weighted estimator is statistically the most efficient \cite{Burgess2013IVW}. In the single exposure setting, when genetic associations with the exposure and outcome are estimated in non-overlapping samples from the same population, weak instruments will bias estimates toward the null \cite{BurgessThompson2011}. However, in the multivariable setting, weak instrument bias may be in any direction, even for two-sample Mendelian randomization \citep{Sanderson2019,Zhu2022}. Thus, weak instrument bias poses a particular barrier in multi-exposure analyses, as effect estimates cannot be conservatively assumed to be biased toward the null as they can be in the single exposure case.
		
		Conditionally weak instruments may occur due to a variety of reasons in multivariable studies. Highly correlated genetic effects on exposures will lead to conditionally weak genetic associations. This may occur, for example, when the exposures are closely related with each other, or when there is correlation among the genetic instruments, such as in cis-Mendelian randomization applications for drug development \cite{Batool2022, Burgess2023}. Measurement error on exposures may also result in weak instruments even if the genetic variants are strong predictors of the true exposures \cite{Zhu2022}. Mediators on the pathway between an exposure and outcome may be strongly associated with genetic predictors of the exposure, but conditionally weakly associated with them. A multivariable Mendelian randomization analysis with mediators as exposures, therefore, will be susceptible to weak instrument bias \cite{Carter2021}.
		
		There have been relatively few methods proposed which effectively account for weak instrument bias in the multivariable Mendelian randomization setting. With individual-level data, the limited information maximum likelihood method is less biased with multiple weak instruments compared with the commonly used two-stage least squares method \cite{BurgessThompson2013, Davies2015}. With summary-level data, a common approach in the single exposure setting has been to develop models based on maximum likelihood estimators which incorporate the uncertainty in the genetic variant-exposure associations. Multivariable versions of such approaches have been proposed by Sanderson et al. \cite{Sanderson2021}, Wang et al. \cite{Wang2021}, and Zhu et al. \cite{Zhu2022}. An alternative approach based on the generalised method of moments has been proposed by Patel et al. \cite{Patel2024}.
		
		Bayesian methods for Mendelian randomization have been less utilised than frequentist methods, despite being shown to perform well across a variety of settings. These include applications in univariable analyses in the presence of pleiotropy \cite{Berzuini2020, Bucur2020, Morrison2020, Cheng2022}, weak instruments \cite{BurgessThompson2012}, and correlated instruments \cite{Shapland2018}. In multivariable analyses, Bayesian model averaging has been proposed for risk factor selection \cite{Zuber2020}. Andrews and Mikusheva have shown that a quasi-Bayesian method outperforms the frequentist generalised method of moments in instrumental variables settings with very weak instruments \cite{Andrews2023}.
		
		The MVMR-Horse method uses a Bayesian framework to perform multivariable Mendelian randomization \cite{GrantBurgess2024}. Motivated by the setting where some genetic variants are valid and some are invalid instruments, the method accounts for pleiotropic effects by applying a horseshore shrinkage prior. The model  also accounts for weak instruments by incorporating the standard errors of the genetic variant-exposure association estimates into the model. Simulation studies demonstrated that the method performed particularly well in comparison with alternative approaches when there were weak instruments with varying levels of pleiotropy. The Bayesian framework therefore offers a promising alternative to existing frequentist methods to account for weak instruments in multivariable studies.
		
		In this paper, we introduce a method that we call MVMR-Pony, which is a version of the MVMR-Horse method without the extension to account for invalid instruments. Focusing on settings with multiple valid genetic instruments, we investigate the relative performance of the MVMR-Pony method against frequentist alternatives across a variety of multivariable Mendelian randomization scenarios with conditionally weak instruments. We demonstrate that MVMR-Pony is able to account for weak instrument bias while retaining nominal type I error rates and coverage levels, with comparable power to frequentist methods.
		
	\section*{Methods}
	
	\subsection*{Data generating process}
	Consider $K$ exposures, denoted $X_{k}, k=1, \ldots, K$, which are potentially causally associated with an outcome, $Y$. The exposures and outcome are assumed to be linearly related to $J$ independent genetic variants, denoted $G_{j}, j=1, \ldots, J$, and unmeasured confounders, represented by a single variable and denoted by $U$, as follows.
	\begin{align}
		X_{k} &= \sum_{j=1}^{J} \beta_{Xjk} G_{j} + U + \varepsilon_{Xk} \label{eq:dgp_X}\\
		Y &= \sum_{k=1}^{K} \theta_{k} X_{k} + U + \varepsilon_{Y} ,\label{eq:dgp_Y}
	\end{align}
	where $\varepsilon_{X1}, \ldots, \varepsilon_{XK}, \varepsilon_{Y}$ are independent error terms, that are also independent of $U$. The parameter $\beta_{Xjk}$ represents the effect of the $j$th genetic variant on the $k$th exposure. The direct causal effect of the $k$th exposure on the outcome is denoted by $\theta_{k}$. Note that this model assumes that the $J$ genetic variants are valid instrumental variables for the set of exposures in a multivariable sense. That is, that there are no pleiotropic effects from the genetic variants to the outcome which do not go via one of the measured exposures. The model as written above does not allow for direct effects between the exposures. This will be relaxed later on when we consider mediation scenarios. It should be further noted that it is implied that the effects of the genetic variants on the exposures, and of the exposures on the outcome, are assumed to be linear with no effect modification.
	
	The methods we consider in our simulation study aim to estimate the direct causal effects using summary statistics of the genetic associations with the exposures and the outcome. In practice, these summary statistics are typically obtained from GWAS using linear regression. We denote by $\hat{\beta}_{Xjk}$ the coefficient from regressing $X_k$ on $G_j$, which is an estimate of $\beta_{Xjk}$. We denote by $\hat{\beta}_{Yj}$ the coefficient from regressing $Y$ on $G_j$. The corresponding standard errors are denoted by $\sigma_{Xjk}$ and $\sigma_{Yj}$, respectively.  We assume that the genetic associations with the exposures are measured in a single sample, and the genetic associations with the outcome in a separate independent sample taken from the same population.
	
	In the model considered above, the outcome $Y$ is a continuous random variable. In practice, the outcome may be binary, in which case summary statistics are computed using logistic regression. Although we do not consider this case in what follows, the methods described are all able to be applied using summary statistics obtained from logistic regression. It should be noted, however, that the non-collapsibility of odds ratios can cause  bias, although some empirical evidence suggests that this bias may not be substantial in multivariable Mendelian randomization settings \cite{Zhu2022}.
	
	\subsection*{Causal effect estimation}
	
	\subsubsection*{The inverse-variance weighted method}
	The inverse-variance weighted method (MVMR-IVW) estimates the direct causal effects using a zero-intercept weighted linear regression of the $\hat{\beta}_{Yj}$ on the $\hat{\beta}_{Xj}$. That is, it fits the model
	\begin{align*}\label{eq:ivw}
		\hat{\beta}_{Yj} = \sum_{k=1}^{K} \theta_{k} \hat{\beta}_{Xjk} + \varepsilon_{j}, \quad j = 1, \ldots, J,
	\end{align*}
	where $\varepsilon_{j}$ is assumed to be normally distributed with mean zero and variance $\sigma_{Yj}^2$. The IVW estimator is thus
	\begin{equation}\label{eq:ivwestimator}
	\left(\sum_{j=1}^{J} \sigma_{Yj}^{-2} \hat{\beta}_{Xj\cdot} \hat{\beta}_{Xj\cdot}' \right)^{-1}\sum_{j=1}^{J} \sigma_{Yj}^{-2} \hat{\beta}_{Yj}\hat{\beta}_{Xj\cdot}.
	\end{equation}
	When the $J$ genetic variants are valid instrumental variables, the IVW method provides the most efficient consistent estimator for the direct causal effects. It is equivalent to two-stage least squares estimation in the individual-level data setting \cite{BurgessThompson2015mvmr, BurgessDudbridge2015mvmr}.
	
	The MVMR-IVW model does not incorporate the uncertainty in the genetic variant-exposure association estimates, the $\sigma_{Xj}$s. That is, it assumes that these associations are measured precisely. In the single exposure setting, this is typically justified by the large samples in which the exposure traits are typically measured in, and because genetic associations with exposures are typically stronger than those with outcomes. However, as discussed above, when the sample sizes are small to moderate, and/or the variation in the exposures explained by the genetic variants is low, then excluding the $\sigma_{Xj}$s can lead to weak instrument bias.
		
	\subsection*{Incorporating the uncertainty in the genetic variant-exposure association estimates}
	
	There have been a number of methods for multivariable Mendelian randomization proposed predicated on modelling the genetic associations with the exposures and outcome as normally distributed random variables, as follows.
	\begin{align*}
		\hat{\beta}_{Yj} &\sim N \left( \theta' \beta_{Xj\cdot}, \sigma_{Yj}^2 \right)\\
		\hat{\beta}_{Xj\cdot} &\sim N \left( \beta_{Xj\cdot}, \Sigma_{Xj} \right) ,
	\end{align*}
	where $\theta$ denotes the $k$-length vector of direct causal effects, $\beta_{Xj\cdot}$ denotes the $k$-length vector of the associations between the $j$th variant and each exposure, and $\hat{\beta}_{Xj\cdot}$ the $k$-length vector of their estimates. This model is justified by the fact that these association estimates are linear regression coefficients taken from typically large samples.
	
	One approach to estimating the direct causal effects is to derive a joint likelihood based on this model and to take as estimates the values of the $\theta_k$s which maximise this likelihood. Methods such as GRAPPLE \cite{Wang2021} and MVMR-MLE \cite{Zhu2022} do this using a profile likelihood. That is, the $\beta_{Xj\cdot}$ parameters are treated as nuisance parameters, and profiled out by first fixing $\theta$. As a result, the estimates of $\theta_{k}$, $k=1, \ldots, K$, are those which minimise the function
	\begin{align}
		\sum_{j=1}^J \frac{\left(\hat{\beta}_{Yj} - \theta' \hat{\beta}_{Xj\cdot}\right)^2}{\sigma_{Yj}^2 + \theta' \Sigma_{Xj} \theta}. \label{eq:mle}
	\end{align}
	Confidence intervals can be derived using asymptotic arguments \cite{Wang2021}. A similar estimator, MVMR-GMM, was arrived at using the generalised method of moments \cite{Patel2023}. This approach minimises the objective function $g\left(\theta\right)'\Omega^{-1}\left(\theta\right) g\left(\theta\right)$, where the $j$th element of the vector $g\left(\theta\right)$ is the moment function $\hat{\beta}_{Yj} - \theta'\hat{\beta}_{Xj\cdot}$, and $\Omega\left(\theta\right)$ is a weighting function. When $\Omega\left(\theta\right)$ is an estimate of the variance of the moment functions, the genetic instruments are independent of each other, and the exposures are assumed independent of each other, this objective function is equivalent to (\ref{eq:mle}). Therefore, in the setting with fully valid instruments, the frequentist methods GRAPPLE, MVMR-MLE, and MVMR-GMM, aim to optimise the same objective function. Their software implementations, however, use different methods for performing this optimisation and, as we demonstrate below in our simulation study, can produce different effect estimates in practice. They also take different approaches to dealing with pleiotropic effects when these are considered.
	
	Wu et al. \cite{Wu2025} have proposed an alternative approach named the spectral-regularised IVW (srivw) estimator. The srivw estimator replaces the $\sum_{j=1}^{J} \sigma_{Yj}^{-2} \hat{\beta}_{Xj\cdot} \hat{\beta}_{Xj\cdot}'$ term in (\ref{eq:ivwestimator}) with
	\[
	\sum_{j=1}^{J} \sigma_{Yj}^{-2} \left(\hat{\beta}_{Xj\cdot} \hat{\beta}_{Xj\cdot}' - \Sigma_{Xj} \right) + \phi \left\lbrace \sum_{j=1}^{J} \sigma_{Yj}^{-2} \left(\hat{\beta}_{Xj\cdot} \hat{\beta}_{Xj\cdot}' - \Sigma_{Xj} \right) \right\rbrace ^{-1}
	\]
	where $\phi$ is a tuning parameter. The estimator incorporates the genetic variant-exposure effect covariances to account for weak instrument bias, then includes the regularisation to stabilise the inversion of the resulting matrix.
	
	For each of these methods, the standard deviation terms $\sigma_{Yj}$, $j=1,\ldots,J$, are assumed known, and taken as the corresponding standard errors from GWAS summary statistics. The covariance matrices, $\Sigma_{Xj}$, $j=1,\ldots,J$, are also assumed known. The diagonal entries of these matrices are taken as the variances of the $j$th genetic variant-exposure association estimates. The off-diagonal entries are not easily estimated from GWAS summary statistics. If the genetic associations with the exposures have been estimated in independent samples, these off-diagonal entries will be zero. It is often the case in practical applications that these off-diagonal entries are assumed to be zero even when the exposures are measured in the same or overlapping samples. This assumption may not be reasonable when the exposures and/or the genetic effects are correlated. One way to estimate these entries is to use the sample correlation between exposures as a proxy for the correlation between the estimates of the genetic associations with those exposures. Thus, an estimate of the $\left(k, l\right)$th entry of $\Sigma_{Xj}$ is
	\begin{equation}
		\sigma_{Xjk} \sigma_{Xjl} \textrm{cor}\left(X_k, X_l \right). \label{eq:SigCov}
	\end{equation}
	This will be a reasonable approximation when the proportion of variance in the exposures explained by the genetic variants is low \cite{Sanderson2021, Grant2022}. Alternatively, there have been methods proposed for estimating these correlations using full GWAS summary statistics, for example by estimating the correlation from non-significant genetic variants \cite{BulikSullivan2015, Ray2018}.
	
	\subsection*{A Bayesian approach}
	We propose here an adapted version of the MVMR-Horse model, removing the aspects of the model which are designed to account for pleiotropic effects. As such, the proposed method uses the same likelihood function as frequentist approaches, but performs estimation in a Bayesian paradigm. Of course, the full MVMR-Horse model may be used in practice if pleiotropy is considered an issue. However, the primary purpose of this paper is to compare the Bayesian approach with commonly used frequentist approaches for mitigating weak instrument bias that do not account for pleiotropic effects. The model we fit is thus
	\begin{align*}
		\hat{\beta}_{Yj} &\sim N \left( \theta' \beta_{Xj\cdot}, \sigma_{Yj}^2 \right) \\
		\hat{\beta}_{Xj\cdot} &\sim N \left( \beta_{Xj\cdot}, \Sigma_{Xj} \right),
	\end{align*}
	with the following prior distributions.
	\begin{align*}
		\beta_{Xj\cdot} &\sim N \left( \mu, V_X \right) \\
		\mu &\sim N \left(0, I_K \right) \\
		V_X &\sim N^{+} \left(0, I_K\right) \\
		\theta_k &\sim N \left( 0, 1\right),
	\end{align*}
	where $N^{+} \left( \cdot, \cdot \right)$ denotes the half-normal distribution.
	
	The priors are chosen to be weakly informative, allowing the data to influence the posterior distributions of the parameters. They do not pre-suppose that the instruments are weakly associated with the exposures, but by placing the mass of the prior distributions around zero, they allow for potentially very small effects. Similar prior distributions were used in the MR-Horse method, which showed good performance in both strong and weak instrument settings \cite{GrantBurgess2024}.

	As with the frequentist methods, we assume that the $\sigma_{Yj}$s and $\Sigma_{Xj}$s are known. Again, they can either be taken to be diagonal matrices with $\left(k, k\right)$th entry $\sigma_{Xjk}^2$, or else the non-diagonal entries can be estimated using, for example, the sample trait correlations. When the full covariance matrices are estimated, we propose using an alternative prior distribution for $\beta_{Xj\cdot}$. For this case, we use a Wishart prior with degrees of freedom $K$ and scale matrix $K$ times the sample covariance matrix of $\hat{\beta}_{Xj\cdot}$.
	
	We call this method MVMR-Pony, akin to a pared down version of the MVMR-Horse method. It is implemented by drawing samples from the posterior distribution of the parameters using JAGS via the rjags package in R. We take 10\,000 samples after a burn-in of 10\,000 iterations. The estimate of $\theta_k$ is the mean of the posterior distribution for this parameter. We compute 95\% credible intervals using the 2.5th and 97.5th percentiles of this distribution. In order to assess convergence, we consider R-hat values, with values close to 1 indicating good convergence to the posterior distribution.
	
	\subsection*{Simulation study setup}
	We consider three general scenarios which lead to conditionally weak instruments in the multivariable setting: correlated genetic effects; measurement error; and mediation. Details of each of these scenarios are below. In each case, we generated two datasets of size $N = 10\,000$ with an outcome and $K=2$ exposures. The genetic associations with the exposures were estimated in one dataset, and with the outcome in the other. The parametrisations of the different scenarios were chosen so that the unconditional F-statistics are between approximately 10 and 20. The conditional F-statistics, on the other hand, range from 3.4 down to 0.5. That is, the scenarios represent cases where instruments are unconditionally strong according to the conventional threshold, but conditionally very weak. The proportion of variation in the exposures explained by the instruments is approximately $12.5\%$ in the primary simulation settings with the highest levels of instrument strength, reflecting typical $R^2$ values in practical settings. Across all scenarios considered, including the supplementary simulations, mean $R^2$ values varied between $2\%$ and $15\%$. We repeated each simulation scenario with $1\,000$ replications.
	
	\subsubsection*{Correlated genetic effects}
	In the correlated genetic effects scenario, we generated data from (\ref{eq:dgp_X}) and (\ref{eq:dgp_Y}) with $\theta_1=0.5$, $\theta_2=0$, $J=60$, $\textrm{var}\left(\varepsilon_{Y}\right)=\textrm{var}\left(\varepsilon_{Xk}\right)=\textrm{var}\left(\varepsilon_{U}\right)=1$, and the genetic variant-exposure associations distributed as
	\begin{equation} \label{eq:geneffect}
		\begin{pmatrix}
			\beta_{Xj1} \\ \beta_{Xj2}
		\end{pmatrix}
		\sim N
		\left(
		\begin{pmatrix}
			0.1 \\ 0.1
		\end{pmatrix} ,
		0.04^2\begin{pmatrix}
			1 & \rho \\ \rho & 1
		\end{pmatrix}
		\right)
	\end{equation}
	The parameter $\rho$ represents the correlation between the genetic effects on $X_1$ and $X_2$. As $\rho$ increases, the conditional strength of the instruments decreases. The $\rho$ parameter was varied between $0$ and $0.8$.
	
	\subsubsection*{Measurement error}
	In the measurement error scenario, we generated data from (\ref{eq:dgp_X}) and (\ref{eq:dgp_Y}) with $\theta_1=0.5$, $\theta_2=0$, $J=60$, $\textrm{var}\left(\varepsilon_{Y}\right)=\textrm{var}\left(\varepsilon_{Xk}\right)=\textrm{var}\left(\varepsilon_{U}\right)=1$, and the genetic variant-exposure associations were distributed as in (\ref{eq:geneffect}) with $\rho=0.2$. That is, with a small amount of correlation among the genetic effects. The genetic association estimates were taken not from a regression of the instruments on $X_1$ and $X_2$, but on $X_1^{*}=X_1 + \zeta_1$ and $X_2^{*}=X_2+\zeta_2$, respectively, where
	\[
		\zeta_1 \sim N \left(0, \nu^2\right),\quad \zeta_2 \sim N \left(0, \nu^2\right).
	\]
	In this scenario, the parameter $\nu$ determines the level of measurement error. As $\nu$ increases, the conditional strength of the instruments decreases. The $\nu$ parameter was varied between $0$ and $4$.
	
	\subsubsection*{Mediation}
	In the mediation scenario, we generated the exposures according to
	\begin{align*}
		X_{1} &= \sum_{j=1}^{J} \beta_{Xj1} G_{j} + U + \varepsilon_{X1} \\
		X_{2} &= \alpha X_1 + \sqrt{1-\alpha^2} \left( \sum_{j=1}^{J} \beta_{Xj2} G_{j} + U + \varepsilon_{X2} \right),
	\end{align*}
	with $J=60$ and $\textrm{var}\left(\varepsilon_{Xk}\right)=\textrm{var}\left(\varepsilon_{U}\right)=1$. The genetic variant-exposure associations were distributed as in (\ref{eq:geneffect}) with $\rho=0.2$. The outcome was generated according to (\ref{eq:dgp_Y})	with $\theta_1=\theta_2=0.5$, and $\textrm{var}\left(\varepsilon_{Y}\right)=1$. In this scenario, $X_1$ represents the primary exposure of interest and its effect on $Y$ is mediated by $X_2$. As $\alpha$ increases, the level of mediation increases. The $\alpha$ parameter was varied between $0$ and $0.8$.
	
	\subsubsection*{Estimation methods}
	In addition to MVMR-IVW, we estimated the direct causal effects, $\theta_1$ and $\theta_2$ using the frequentist methods MVMR-GMM, GRAPPLE, and srivw. Because we are primarily interested in the case where all genetic variants are valid instruments, we do not use the default settings for the MVMR-GMM and GRAPPLE methods: we implemented MVMR-GMM without accounting for overdispersion heterogeneity; we implemented GRAPPLE with an $\ell_2$ loss function and the pleiotropy parameter fixed at zero. We further used the MVMR-Pony method described above as a Bayesian approach. For MVMR-GMM, GRAPPLE, srivw, and MVMR-Pony, we applied the methods in two ways: one with the $\Sigma_{Xj}$ matrices input with the off-diagonal entries set to zero, and one with the full $\Sigma_{Xj}$ matrices estimated according to (\ref{eq:SigCov}) using the sample correlation between $X_1$ and $X_2$.
	
	The methods were evaluated in terms of mean estimates, mean bias, coverage, and rejection rates in each scenario and value of $\rho$ (measure of collinearity between genetic effects in Scenario 1), $\nu$ (extent of measurement error in Scenario 2), or $\alpha$ (level of mediation in Scenario 3). Bias is defined as the absolute value of the difference between the estimated and true causal effects. Coverage is defined as the proportion of replications for which the 95\% confidence interval or 95\% credible interval includes the true value of $\theta_k$. The rejection rate is the proportion of replications for which zero is not contained within the 95\% confidence interval or 95\% credible interval. For Scenarios 1 and 2, the rejection rate for $\theta_1$ represents power, and for $\theta_2$ the type I error rate. For Scenario 3, the rejection rate represents power for both $\theta_1$ and $\theta_2$. The unconditional F-statistics were computed as the mean squared t-statistic for each instrument, and the conditional F-statistics were computed using the method implemented in the MendelianRandomization R package \cite{Patel2023}.
	
	\subsection*{Supplementary simulation settings}
	In addition to the primary simulation scenarios, we considered five supplementary scenarios, where the data were generated as above but with the following differences.
	\begin{enumerate}[label=S\arabic*, start=1]
		\item The mean genetic effects on $X_1$ and $X_2$ had opposite signs. That is, $E\left(\beta_{Xj1}\right) = -E\left(\beta_{Xj2}\right) = 0.1$.
		\item The genetic effects on $X_1$ and $X_2$ were negatively correlated. That is, the $\rho$ parameters had the same magnitude but were negative.
		\item The genetic effects on $X_1$ and $X_2$ had opposite signs ($E\left(\beta_{Xj1}\right) = -E\left(\beta_{Xj2}\right) = 0.1$) and were also negatively correlated ($\rho<0$)
		\item The strength of the genetic associations with the causal exposure $X_1$ was increased, with $E\left(\beta_{Xj1}\right) = 0.2$, and the genetic associations with the null exposure $X_2$ was decreased, with $E\left(\beta_{Xj1}\right) = 0.05$ (note, for the mediation scenario, both $X_1$ and $X_2$ are causal exposures).
		\item The strength of the genetic associations with the causal exposure $X_1$ was decreased, with $E\left(\beta_{Xj2}\right) = 0.2$, and the genetic associations with the null exposure $X_2$ was increased, with $E\left(\beta_{Xj1}\right) = 0.05$ (note, for the mediation scenario, both $X_1$ and $X_2$ are causal exposures).
	\end{enumerate}
	As in the primary simulation study, in the supplementary settings the genetic instruments are unconditionally strong, with F-statistics above 10 in most cases (the exception being in the mediation scenarios, where some unconditional F-statistics relating to the mediator are low, with a minimum value of 2.7). The conditional F-statistics, on the other hand, range from 12.0 down to 0.4, and are below 10 in all but two cases.

	\section*{Results}
	
	\subsection*{Primary simulation scenario}
\begin{figure}[p]
		\thisfloatpagestyle{empty}
		\centering
		\includegraphics{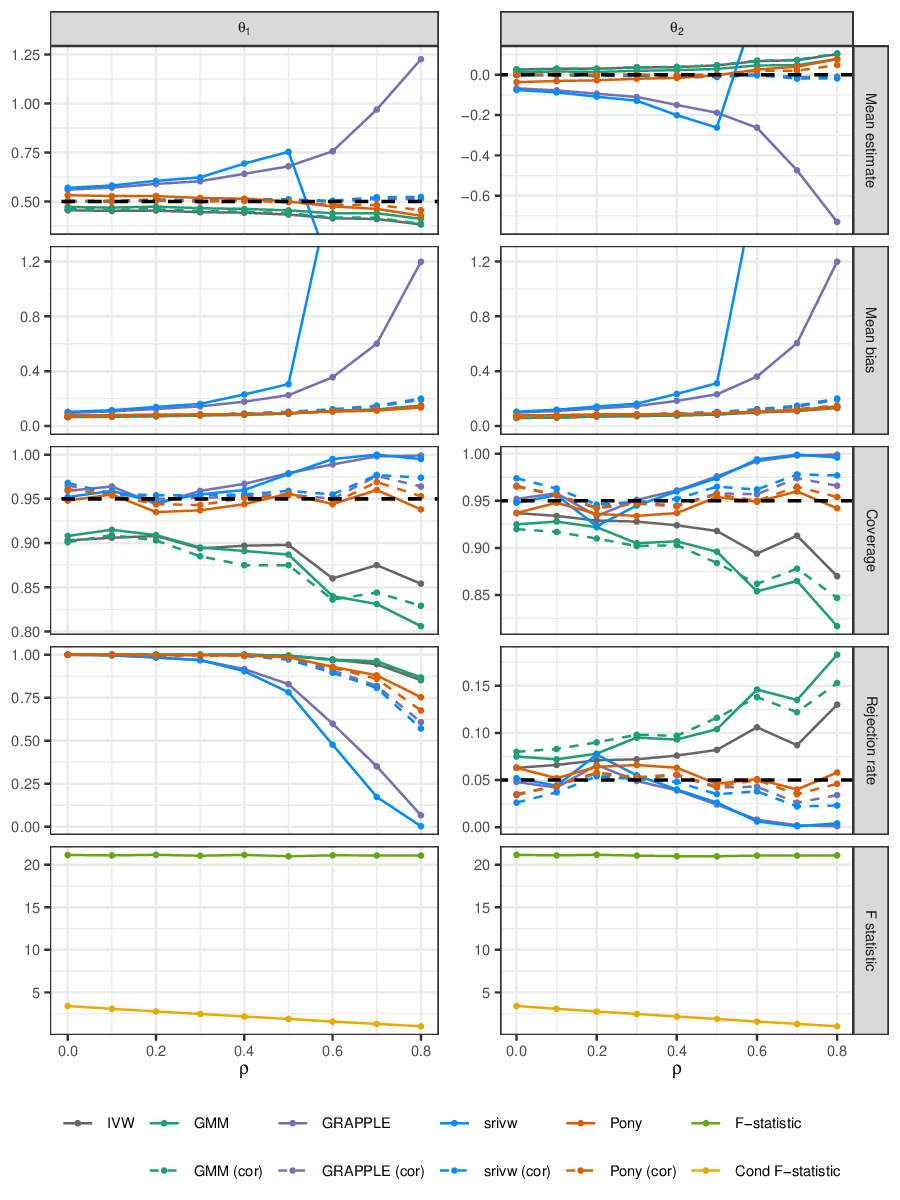}
		\caption{Simulation results for the scenario with increasingly correlated genetic effects, $\rho$, showing the mean estimate, coverage, and rejection rate (that is, power) using the IVW, GMM, GRAPPLE, srivw, and MVMR-Pony methods. The dashed lines indicate the true causal effect (for mean estimate) or the nominal level (for coverage). Also shown are the F statistics and conditional F statistics for each value of $\rho$. Note that some values for the srivw method are beyond the displayed range.}
		\label{fg:S1}
	\end{figure}

	Figure \ref{fg:S1} shows the results for the scenario with correlated genetic effects. As the level of correlation, $\rho$, increases, the conditional F-statistics decrease while the unconditional F-statistics remain constant. GRAPPLE and srivw have considerable bias in the moderate to high levels of correlation. The bias from the MVMR-Pony method is similar to all other methods. MVMR-Pony also retains coverage and type I error rates at nominal levels, whereas MVMR-IVW and MVMR-GMM have lower power and higher type I error rates. When the full $\Sigma_{Xj}$ matrices are estimated and input into the models, the bias in the GRAPPLE and srivw estimators reduce and are similar to MVMR-Pony, although the mean bias remains higher for both, particularly in the higher genetic correlation settings. Coverage and type I error rates are at nominal levels for GRAPPLE and srivw when full $\Sigma_{Xj}$ matrices are used. In contrast, MVMR-Pony gives good results even when this matrix is not provided.
	
\begin{figure}[p]
		\thisfloatpagestyle{empty}
		\centering
		\includegraphics{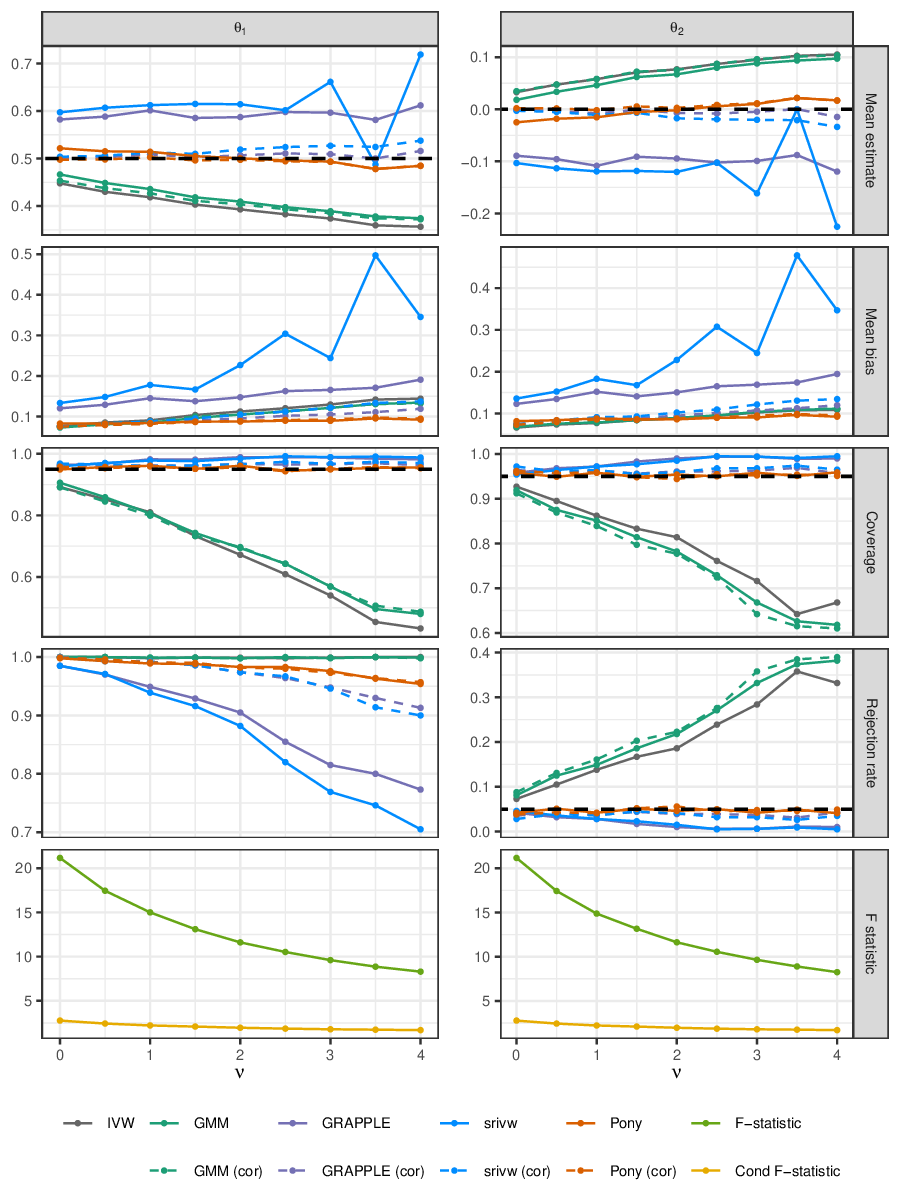}
		\caption{Simulation results for the scenario with increasing measurement error on exposures, $\nu$, showing the mean estimate, coverage, and rejection rate (that is, power) using the IVW, GMM, GRAPPLE, srivw, and MVMR-Pony methods. The dashed lines indicate the true causal effect (for mean estimate) or the nominal level (for coverage). Also shown are the F statistics and conditional F statistics for each value of $\nu$.}
		\label{fg:S2}
	\end{figure}

	Figure \ref{fg:S2} shows the results for the scenario with measurement error. Here, both conditional and unconditional F-statistics reduce as the level of measurement error, $\nu$, increases. Nonetheless, the mean unconditional F-statistics only drop below $10$ at the highest levels of $\nu$. As before, GRAPPLE and srivw show considerable bias when the full $\Sigma_{Xj}$ matrices are not estimated, and this improves substantially when they are. With no or low levels of measurement error, the mean bias is lowest for MVMR-GMM, however as measurement error increasses, the mean bias from MVMR-Pony remains at relatively low levels while it increases for all other methods. MVMR-Pony also has higher power than GRAPPLE and srivw, as well as type I error rates closer to the nominal level. As before, MVMR-Pony is not sensitive to misspecification of the trait covariance matrix.
	
\begin{figure}[p]
			\thisfloatpagestyle{empty}
			\centering
			\includegraphics{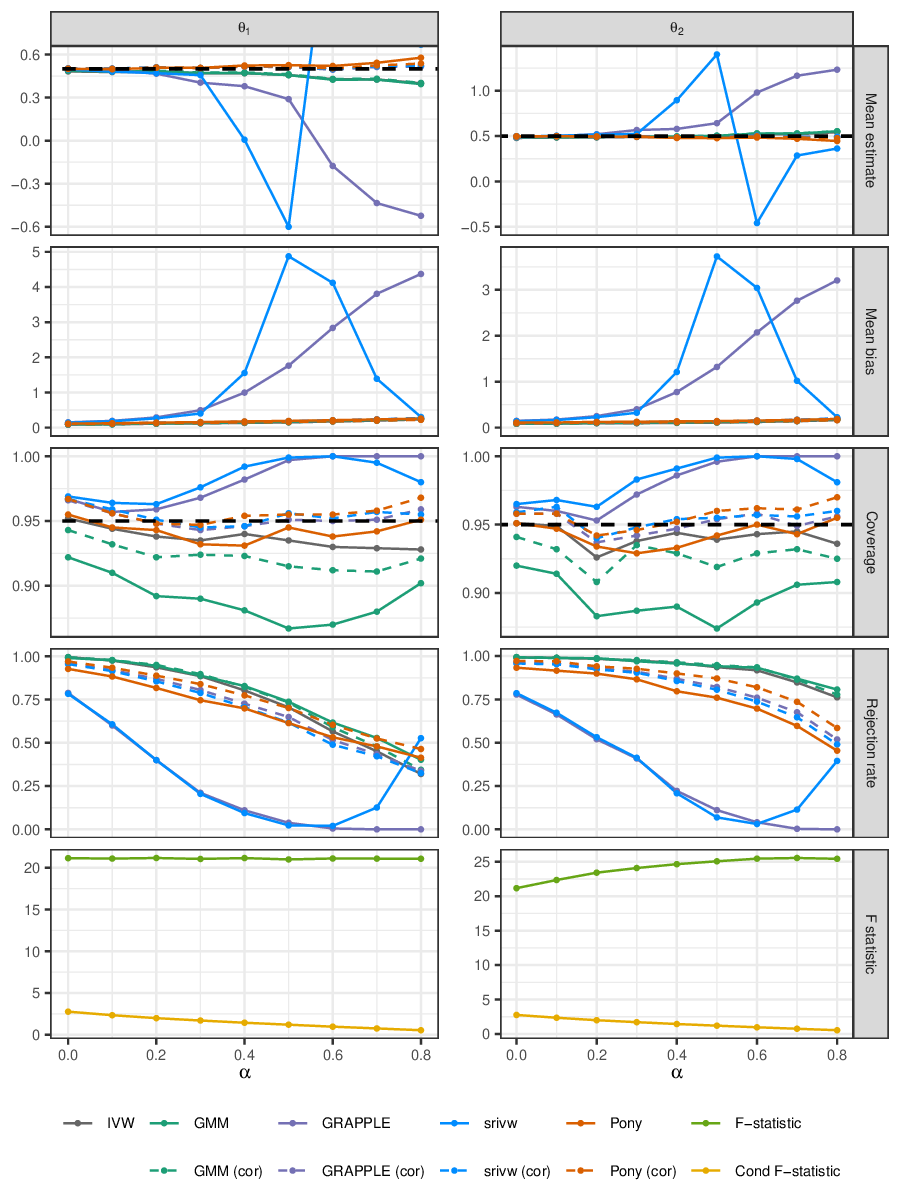}
			\caption{Simulation results for the scenario with increasing level of mediation, $\alpha$, showing the mean estimate, coverage, and rejection rate (that is, power) using the IVW, GMM, GRAPPLE, srivw, and MVMR-Pony methods. The dashed lines indicate the true causal effect (for mean estimate) or the nominal level (for coverage). Also shown are the F statistics and conditional F statistics for each value of $\alpha$. Note that some values for the srivw method are beyond the displayed range.}
			\label{fg:S3}
		\end{figure}

	Figure \ref{fg:S3} shows the results for the mediation scenario. Here, the unconditional F-statistics decrease as the proportion of the effect mediated increases. The unconditional F-statistics remain constant for $X_1$ and increase for $X_2$. The pattern of the results is similar to the previous scenarios. GRAPPLE and srivw are substantially biased in the conditionally weaker settings, with this bias reducing when full $\Sigma_{Xj}$ matrices are used. MVMR-IVW, MVMR-GMM, and MVMR-Pony have similar levels of mean bias, but MVMR-Pony has coverage much closer to nominal levels. MVMR-Pony generally has higher power than GRAPPLE and srivw, although MVMR-IVW and MVMR-GMM have the highest power in this scenario.
	
	\subsection*{Supplementary simulations}
	The results from the supplementary simulations with correlated genetic effects are shown in Figure \ref{fg:S1_th1_supp} (showing the results for estimates of $\theta_1$) and in Figure \ref{fg:S1_th2_supp} ($\theta_2$). Similarly, the results for the measurement error scenarios are shown in Figures \ref{fg:S2_th1_supp} and \ref{fg:S2_th2_supp}, and for the mediation scenarios in Figures \ref{fg:S3_th1_supp} and \ref{fg:S3_th2_supp}.
	
	Consistently across all the supplementary simulations, the mean bias from MVMR-Pony is comparable or lower than all other methods. The coverage and type I error rates at or close to nominal levels in all settings, which is not the case for any other method considered. The only settings where MVMR-Pony has coverage and type I error rates substantially away from the nominal levels is those with very high negative genetic correlation (supplementary scenarios S2 and S3), and even then only when the full $\Sigma_{Xj}$ matrices are not used. GRAPPLE and srivw perform similar to MVMR-Pony in many cases. However, particularly at very low conditional instrument strength settings, these methods do not do well without the full $\Sigma_{Xj}$ matrices being estimated. In contrast, MVMR-Pony typically performs much better across all metrics considered when these matrices are not used.
		
	\section*{Applied example}
	
	Reduced kidney function has been linked with a range of cardiometabolic diseases including hypertension \cite{Staplin2022}, cardiovascular disease \cite{Gaziano2022}, and dyslipodemia \cite{Bulbul2018}. Dobrijevic et al. considered the causal effects of kidney disease on cancer outcomes, using as exposures estimated glomerular filtration rate (eGFR) and urinary albumin-creatine ratio (UACR)\cite{Dobrijevic2024}. In a multivariable analysis, they selected 67 independent ($r^2<0.001$) genetic variants associated with either exposure at genome-wide significance. The genetic association estimates with eGFR were taken from the GWAS of \citet{Wuttke2019}, including $567\,460$ individuals of European ancestry, and with UACR from the GWAS of \citet{Teumer2019}, including $127\,865$ individuals predominantly of European ancestry. To mitigate against horizontal pleiotropy, this set of instruments excluded genetic variants which were associated with selected confounders including type 2 diabetes, body mass index, smoking, hypertension, income, and education. Genetic variants associated with eGFR were also excluded if they also associated inversely with blood urea nitrogen at a Bonferroni corrected level. Estimating the unconditional F-statistic as the mean of the squared t-statistic for each variant, this set of instruments are unconditionally strong for each exposure, with average F-statistics of 75.0 and 10.3 for eGFR and UACR, respectively. The conditional F-statistics for the exposures are 8.6 and 8.5, suggesting there may be some weak instrument bias in the multivariable setting.
	
	Here, we consider the conditional associations of genetically-predicted eGFR and UACR with Alzheimer's disease using the same set of genetic instruments as \citet{Dobrijevic2024}. Genetic associations with Alzheimer's disease are taken from the GWAS of \citet{Lambert2013}, which included $74\,046$ individuals of European ancestry. Three variants were not in the outcome dataset and so were not included in the analysis. We performed univariable and multivariable IVW estimation, and also implemented MVMR-GMM, GRAPPLE, and MVMR-Pony, assuming there is no bias from pleiotropy. We repeated the analysis using pleiotropy-robust versions of these approaches, that is, the univariable and multivariable Median methods \cite{Bowden2016median, GrantBurgess2021mvmr}, MVMR-GMM accounting for overdispersion heterogeneity, GRAPPLE with the robust loss function and pleiotropy parameter activated, and MVMR-Horse.
	
\begin{figure}[p]
	\thisfloatpagestyle{empty}
	\centering
	\includegraphics{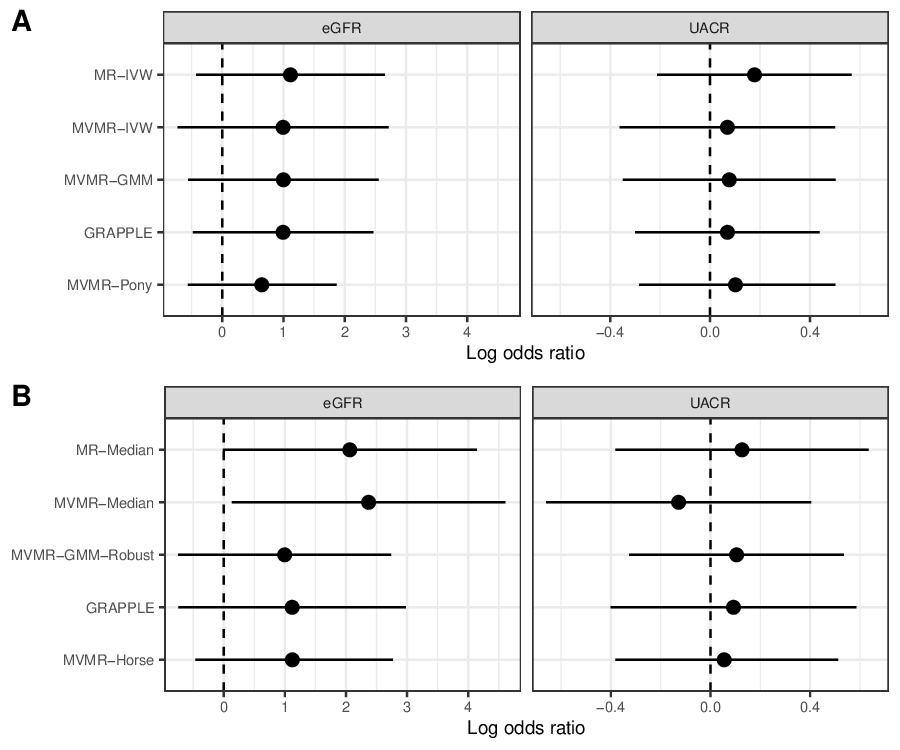}
	\caption{Log odds ratio for Alzheimer's disease point estimate and 95\% confidence interval (or credible interval) per unit increase in eGFR and UACR, using (A) non-pleiotropy-robust methods, and (B) pleiotropy-robust methods.}
	\label{fg:appliedex}
\end{figure}

	The results are shown in Figure \ref{fg:appliedex}. Both univariable and multivariable IVW estimation suggests there is no causal effect of either exposure on the risk of Alzheimer's disease. The weak instrument robust methods support this finding. Causal effect estimates for both exposures using GMM and GRAPPLE are very close to that of MVMR-IVW, with the MVMR-Pony estimate attenuating slightly toward the null for eGFR, and slightly away from the null for UACR. The pattern of results are similar when pleiotropy-robust methods are used.
	
	We note that the Bayesian methods, MVMR-Pony and MVMR-Horse, had R-hat values of $1.0$ in all cases. Although the Bayesian methods had longer run-times compared with the frequentist methods, this was not prohibitive. MVMR-Pony ran in 3 seconds, and the full MVMR-Horse method completed in under 2 minutes, on a machine running an Apple M1 Pro Chip.
	
	\section*{Discussion}
	
	In this paper, we have introduced MVMR-Pony, a Bayesian method for multivariable Mendelian randomization that gives accurate estimates and inferences with weak instruments. Our simulation study has demonstrated that the MVMR-Pony method outperforms existing frequentist methods across various metrics in all of the weak instrument bias scenarios considered. In particular, the approach was able to estimate causal effects with relatively low levels of bias while retaining nominal coverage and type I error rates. It also maintained comparable power to the alternative methods. It further demonstrates the importance of considering conditional instrument strength in multivariable Mendelian randomization studies. In most of the settings considered, the genetic instruments would be considered unconditionally strong instruments for both exposures according to the conventional F-statistic threshold.
	
	The maximum likelihood method, GRAPPLE, appeared to have difficulty in converging to the true solution, particularly in settings with very weak conditional instrument strength. The srivw method had similar difficulties, despite its use of regularisation to stabilise the estimator. On the other hand, the Bayesian approach, MVMR-Pony, was able to converge to the true posterior distributions, evidenced by mean R-hat values which were below $1.005$ for both causal effect parameters in all cases. When the full covariance matrices of the genetic association estimates were incorporated into the model, GRAPPLE and srivw both performed much better. Nonetheless, MVMR-Pony retained comparable or superior performance. Furthermore, estimates of these covariance matrices are not always easily obtained in practice. They require accurate estimates of the sample trait correlation which is not always available in the summary-level settings.
	
	In terms of coverage, power, and type I error rates, MVMR-Pony typically showed as good or better results compared with the other methods considered, albeit not uniformally across all scenarios. This suggests that the Bayesian framework used by the method provides better quantification of the uncertainty in the causal effect estimates, which is in line with previous work looking at Bayesian methods for instrumental variables estimation \cite{BurgessThompson2012}.
		
	Not considered here are settings with invalid genetic instruments due to genetic pleiotropy. The frequentist methods, GRAPPLE, MVMR-GMM, and srivw, can all be extended to account for balanced pleiotropy. In the case of the former, a robust loss function can be applied and a heterogeneity parameter included. The MVMR-GMM and srivw methods can also allow for overdispersion to account for pleiotropy. In the case of MVMR-Pony, the full MVMR-Horse model can be used to account for directional, and potentially also correlated, pleiotropy. An alternative maximum likelihood-based method, MVMR-cML, can also account for directional and correlated pleiotropy \cite{Xue2023}. However, previous simulation studies have suggested that the MVMR-Horse method performs particularly well in comparison with MVMR-cML in a setting with conditionally weak instruments induced by a relatively low sample size and high error variance on the exposures \cite{GrantBurgess2024}.
	
	It should also be noted that the simulation settings here consider scenarios with very low conditional instrument strength. In settings where conditional F-statistics are closer to the typical threshold of $10$, the frequentist methods considered would be expected to have improved performance, and likely comparable performance with the Bayesian MVMR-Pony method. Nonetheless, practical applications with very low conditional instrument strength are increasingly common.
	
	Overall, we have shown that the proposed MVMR-Pony method for multivariable Mendelian randomization provides an effective and flexible approach for causal effect estimation and can provide valid inference in the presence of conditionally weak instruments. It is an important tool for Mendelian randomization given the increasing emergence of multi-exposure applications. It is of particular importance for practitioners working in settings with very low conditional instrument strength, where existing frequentist methods are not able to provide reliable inference.
	
	\section*{Data and code availability}
	R code for performing the MVMR-Pony method, and for reproducing the simulation results and applied analyses, can be found at \url{https://github.aj-grant/mvmrpony}. The code makes use of the packages MendelianRandomization \cite{Yavorska2017, Broadbent2020}, GRAPPLE \cite{GRAPPLEpackage}, mr.divw \cite{Wu2025}, mrhorse \cite{GrantBurgess2024}, rjags \cite{rjagspackage}, and R2jags \cite{R2jagspackage}. Data used in the applied example are publicly available and can be accessed via the references given.
	
	\section*{Acknowledgements}
	SB is supported by the Wellcome Trust (225790/Z/22/Z).

	\bibliographystyle{unsrtnat_ag}
	\bibliography{mvmr_weak_instruments_bib}
	
	\renewcommand{\thefigure}{S\arabic{figure}}
	\setcounter{figure}{0}
	\newpage
	\newgeometry{bottom=2.2cm}
	\begin{landscape}
		\begin{figure}[p]
			\thisfloatpagestyle{empty}
			\centering
			\includegraphics[]{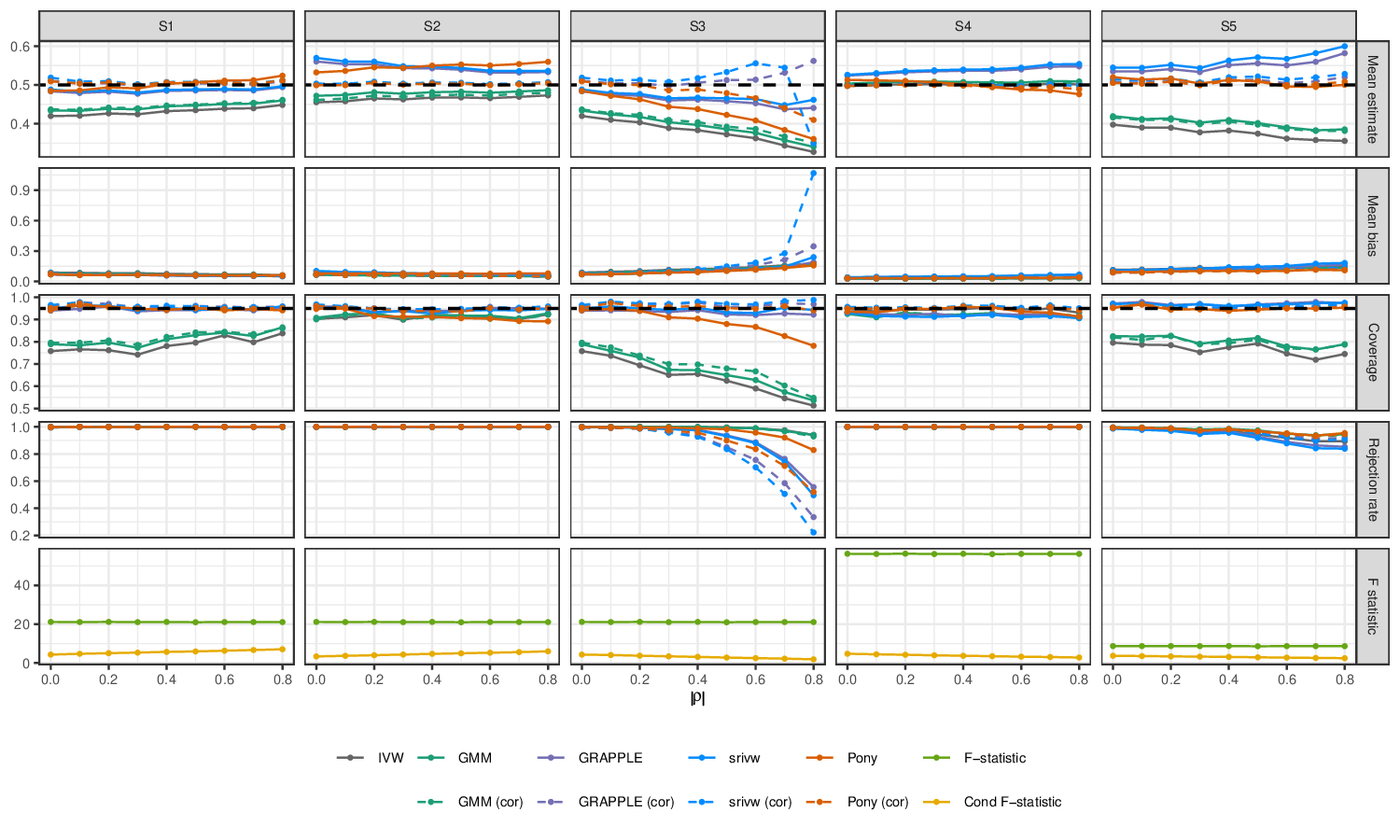}
			\caption{Simulation results for the estimation of $\theta_1$ in the supplementary scenarios with increasing correlated genetic effects.}
			\label{fg:S1_th1_supp}
		\end{figure}
		
		\begin{figure}[p]
			\thisfloatpagestyle{empty}
			\centering
			\includegraphics[]{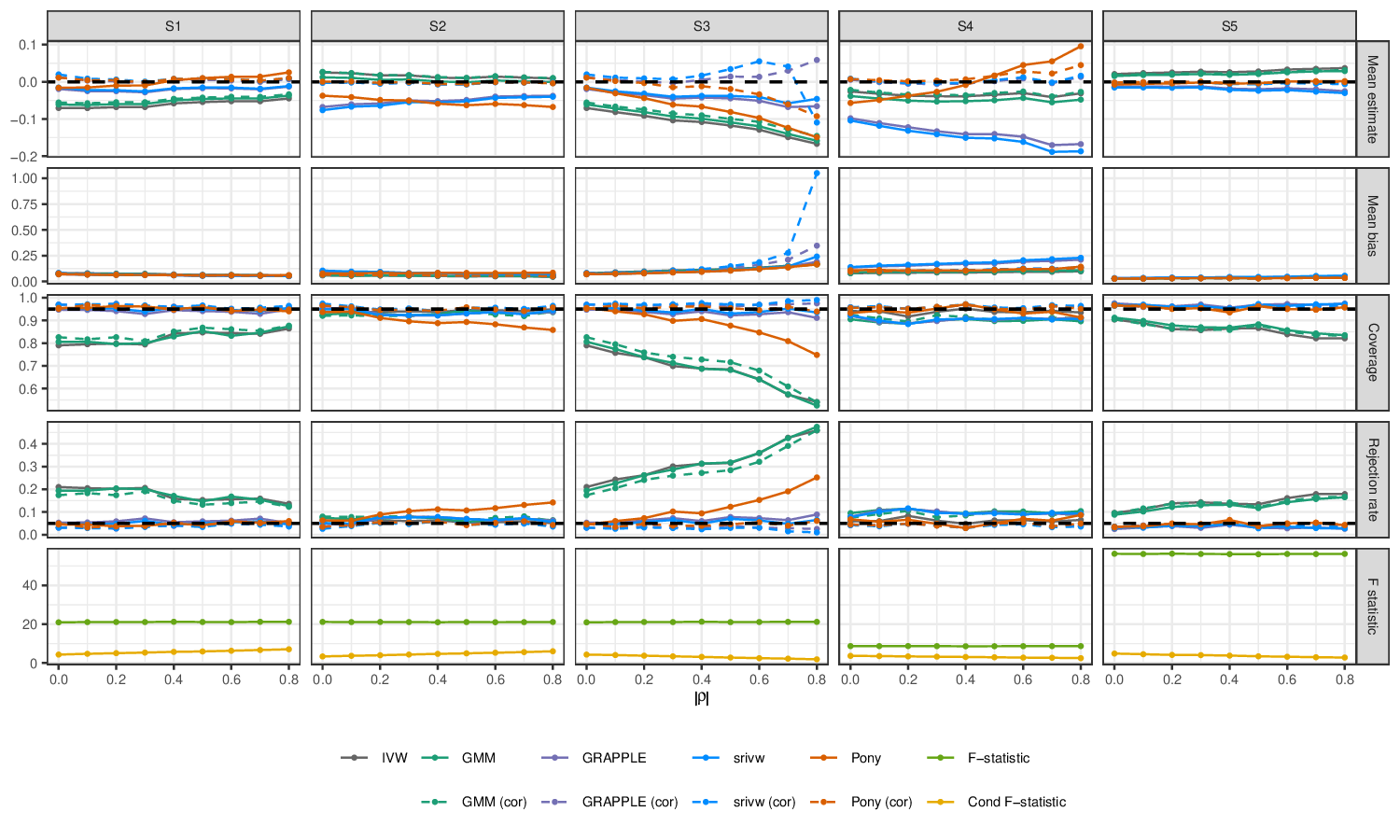}
			\caption{Simulation results for the estimation of $\theta_2$ in the supplementary scenarios with increasing correlated genetic effects.}
			\label{fg:S1_th2_supp}
		\end{figure}
		
		\begin{figure}[p]
			\thisfloatpagestyle{empty}
			\centering
			\includegraphics[]{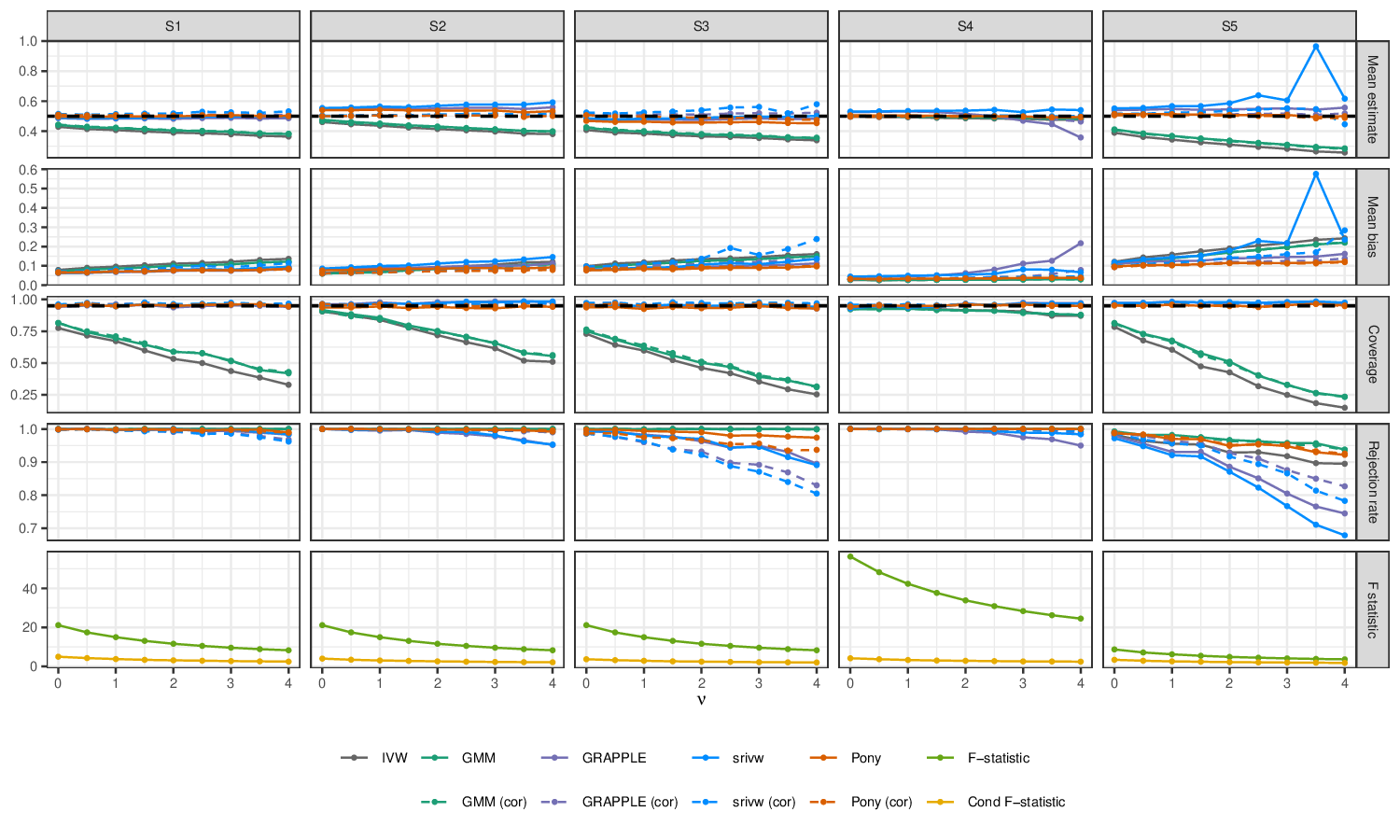}
			\caption{Simulation results for the estimation of $\theta_1$ in the supplementary scenarios with increasing measurement error.}
			\label{fg:S2_th1_supp}
		\end{figure}
		
		\begin{figure}[p]
			\thisfloatpagestyle{empty}
			\centering
			\includegraphics[]{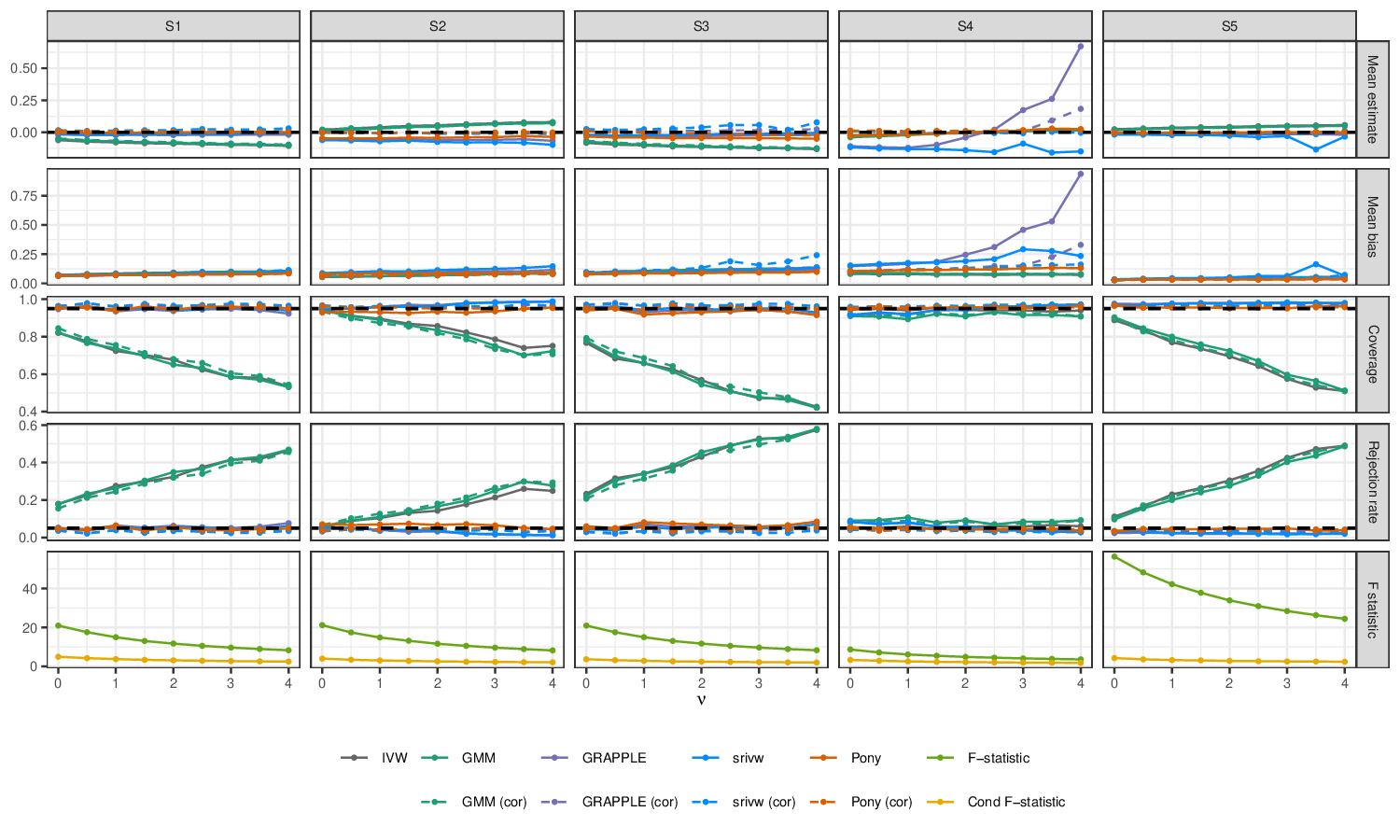}
			\caption{Simulation results for the estimation of $\theta_2$ in the supplementary scenarios with increasing measurement error.}
			\label{fg:S2_th2_supp}
		\end{figure}
		
		\begin{figure}[p]
			\thisfloatpagestyle{empty}
			\centering
			\includegraphics[]{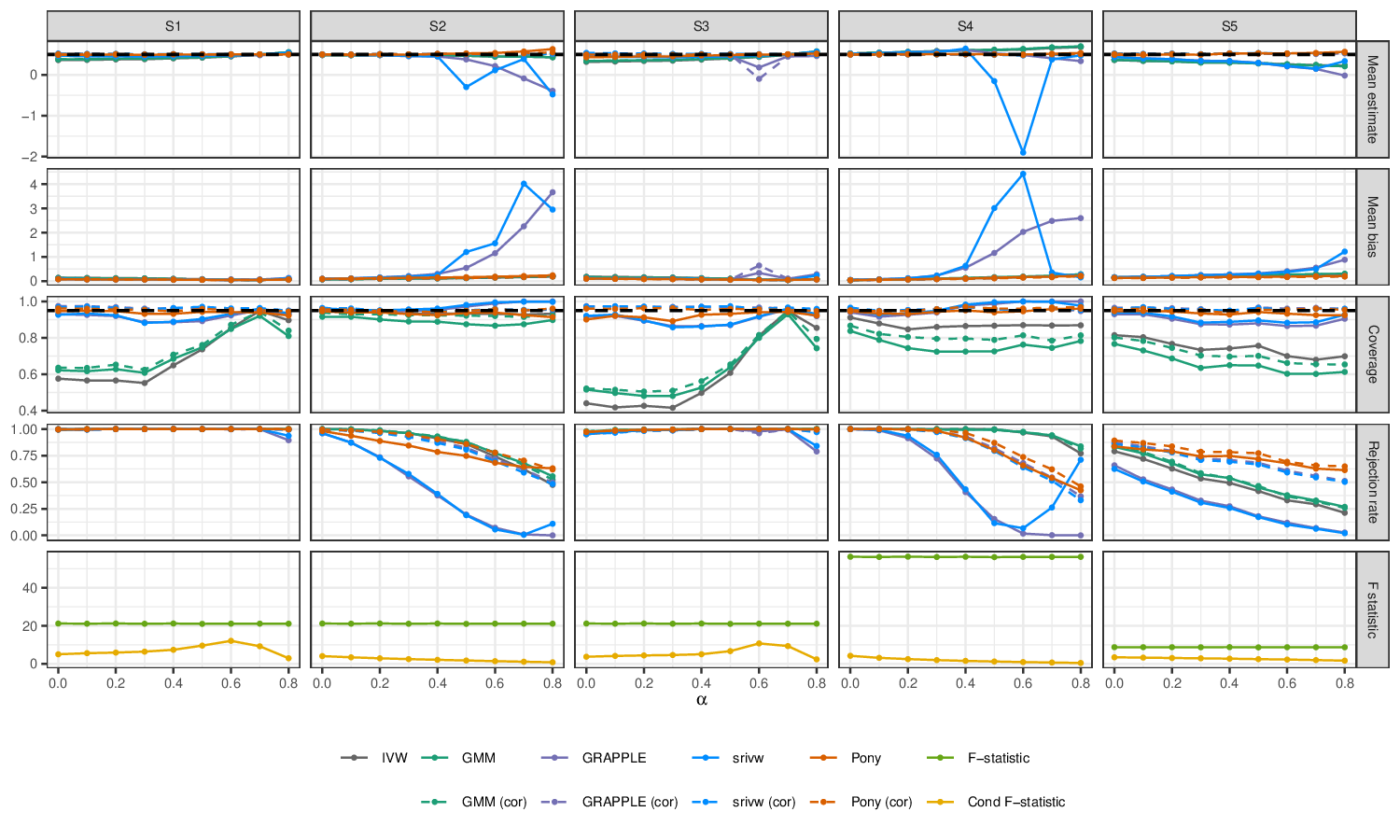}
			\caption{Simulation results for the estimation of $\theta_1$ in the supplementary scenarios with increasing level of mediation.}
			\label{fg:S3_th1_supp}
		\end{figure}
		
		\begin{figure}[p]
			\thisfloatpagestyle{empty}
			\centering
			\includegraphics[]{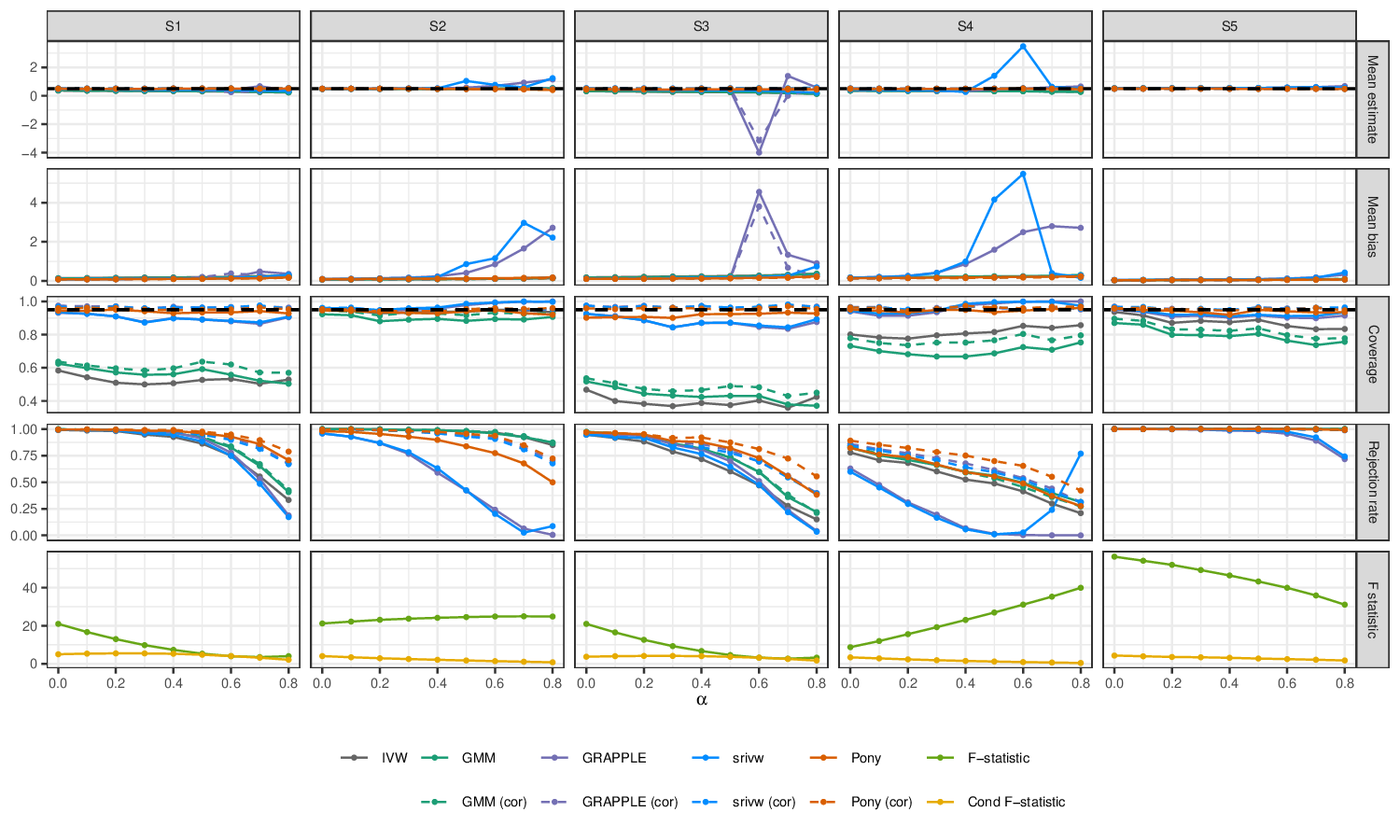}
			\caption{Simulation results for the estimation of $\theta_2$ in the supplementary scenarios with increasing level of mediation.}
			\label{fg:S3_th2_supp}
		\end{figure}
		
	\end{landscape}
	\restoregeometry

\end{document}